
\input phyzzx

\def\PS{{\Gamma}}
\def\Real{{\bf R}}
\def\Spin{{\bf S}}
\def\hidV{{\bf U}}
\def\ez{{\bf e}_z}
\def\Rot{{T_{\theta}}}
\def\Rotz{{R_{z}}}
\def\Roty{{R_{y}}}

\def\Dy{{{\mit \Delta}y}}

\def\TA{{\tau_A}}
\def\TB{{\tau_B}}
\REF\NoGo{J. S. Bell, Rev. Mod. Phys. ${\bf 38}$ (1966), 447-52;
Kochen, S., E.P. Specker, Journ. Math. Mech. ${\bf 17}$ (1967), 59-87}
\REF\BelCas{ E. G. Beltrametti, G. Cassinel, {\it The Logic of Quantum
Mechanics} (Addison-Wesley, 1981),
ENCYCLOPEDIA OF MATHEMATICS and Its Applications, vol 15, Chap. 15
}
\REF\EPRB{ J. S. Bell, Physics {\bf 1} (1964), 195-200}
\REF\EXP{S. J. Freedman, J. F. Clauser, Phys. Rev. Lett. {\bf 28} (1972),
938-41;
E. S. Fry, R. C. Thompson, Phys. Rev. Lett. {\bf 37} (1976), 465-68;
M. Lamehi-Rachti, W. Mittig, Phys. Rev. {\bf D14} (1976), 2543-55;
A. Aspect, P. Grangier, G. Roger, Phys. Rev. Lett. {\bf 47} (1981), 460-3,
Phys. Rev. Lett. {\bf 49} (1982), 91-4, Phys. Rev. Lett. {\bf 49} (1982),
1804-7;
W. Perrie, A. J. Duncan, H. J. Beyer, H. Kleinpoppen, Phys. Rev. Lett. {\bf 54}
(1985), 1790-3}
\REF\CH{J. F. Clauser, M. A. Horne, Phys. Rev. {\bf D10} (1974), 526-35}
\REF\LOOPH{T. W. Marshall, E. Santos, F. Selleri, Phys. Lett. {\bf 98A} (1983),
5-9}
\REF\Sca{G. C. Scalera, Lett. Nuovo Cimento {\bf 38} (1983), 16-8, Lett. Nuovo
Cimento {\bf 40} (1984), 353-61}
\REF\Not{S. Notarrigo, Nuovo Cimento {\bf 83B} (1984), 173-87}
\REF\Pas{S. Pascazio, Phys. Lett. {\bf 118A} (1986), 47-53}
\REF\Bohm{D. Bohm, {\it Quantum Theory} (Prentice-Hall, Inc., New York, 1951),
Chap. 22.}
\REF\ATTR{ Roughly speaking, an attractor is an invariant set on which
neighboring trajectories accumulate. In our case, $\alpha_1$ and $\alpha_2$ are
just limit cycles.
See
J. -P. Eckmann, D. Ruelle, Rev. Mod. Phys. {\bf 57}(1985), 617-56.}
\REF\Redh{M. Redhead, {\it Incompleteness, Nonlocality, and Realism} (Clarendon
Press, Oxford, 1987), $\S$ 4.1. p. 85.}
\REF\Asp{A. Aspect, P. Grangier, G. Roger, Phys. Rev. Lett. {\bf 49} (1982),
91-4}
\REF\Watanabe{S. Watanabe, {\it KNOWING AND GUESSING - A Quantative Study of
Inference and Information} (Wiley, 1969)}

\FIG\PROBfig{
 The probability of $S_z$ being $+j$ for the ensemble of states distributed
uniformly in $\beta_\theta$ plotted versus the relative angle $\theta$.
$\diamond$ represents the results of our model and the solid curve is the
corresponding results of quantum mechanics.}
\FIG\NONCONfig{
 The correlation  of $\Spin_A\cdot \ez$  and $\Spin_B\cdot \Rot \ez$ for the
ensemble of states distributed uniformly in $s$ plotted versus the relative
angle $\theta$.
$\diamond$ represents the results of our model and the solid curve is the
corresponding results of quantum mechanics.
The value of the closing time in sampling $T$ is 0.133.
}
\FIG\CLOfig{
 The correlation  of $\Spin_A\cdot \ez$  and $\Spin_B\cdot \Rot \ez$ without
the closing time for the ensemble of states distributed uniformly in $s$
plotted versus the relative angle $\theta$.
$\diamond$ represents the results of our model and the solid curve is the
corresponding results of quantum mechanics.
}
\FIG\Ffig{
Graph of $F(\phi)$ given in ref. $(\Redh)$ against the relative angle $\phi$.
$\diamond$ represents the results of our model with the closing time $T$=0.133.
$\times$ represents the results of our model without the closing time.
The solid curve is the corresponding results of quantum mechanics.}
\FIG\Devices{
A schematic diagram of the arrangement of the devices.
Each clock shows the time when the object whose finishing time is less than $T$
exists at each indicated place.}
\FIG\SAMPfig{
Graph of the number of samples in the calculation of the correlations for the
ensemble of states distributed uniformly in $s$ against the relative angle.
The value of the closing time $T$ is 0.133.}

\Pubnum={EPHOU-94-006}
\pubnum={ }
\titlepage
\title{ Cannot Local Reality Exist in the EPR-Bohm Gedankenexperiment ?   }
\author{ Satoshi Uchiyama }
\address{ Department of Physics, Faculty of Science,\break
 Hokkaido University, Sapporo 060, Japan}
\address{{\rm (e-mail: uchiyama@particle.phys.hokudai.ac.jp)}}
\abstract{
We model measuring processes of a single spin-1/2 object and of a pair of
spin-1/2 objects in the EPR-Bohm state by  systems of differential equations.
Our model is a local model with hidden-variables of the EPR-Bohm
{Gedankenexperiment}.
Although there is no dynamical interaction between a pair of spin-1/2 objects,
the model can reproduce approximately the quantum-mechanical correlations by
the coincidence counting.
Hence the Bell inequality is violated.
This result supports the idea that the coincidence counting is the source of
the non-locality in the EPR-Bohm {Gedankenexperiment}.
}
\endpage			

\chapter{Introduction}

It is known that quantum mechanics can not be reduced to any non-contextual
hidden-variable theories$^{(\NoGo)}$.
This means that the probability space of outcomes of measurements changes
according to what is measured.
A hidden-variable theory with such many probability spaces is often called a
{\it contextual} hidden-variable theory$^{(\BelCas)}$.
In such a theory, we usually consider that the change of the probability space
is due to the interaction between the object and the measuring apparatus.
Bell argues that in the EPR-Bohm {Gedankenexperiment}, this interpretation
leads us to an unacceptable conclusion.
He insists that if the Bell inequality is not satisfied, then there exists
action at a distance in the EPR-Bohm {Gedankenexperiment}$^{(\EPRB)}$.
Several EPR-Bohm type experiments have already been performed since then and
violations of the Bell type inequalities have been observed$^{(\EXP)}$.
As a result, it has been widely believed that quantum mechanics has a non-local
character such as action at a distance.
It is hardly known, however, that  several authors$^{(\CH - \Pas)}$ showed that
violation of the Bell inequality did not always mean  existence of the action
at a distance, making  local models that violate the Bell type inequalities,
about ten years ago.
Especially,  Scarela$^{(\Sca)}$, Notarigo$^{(\Not)}$ and Pascazio$^{(\Pas)}$
argue that the coincidence counting is a source of the non-locality.
They make only  models for photons, since most of the experiments were
performed for pairs of photons.

In this paper, we shall present a local model for spin-1/2 objects, in
accordance with the Bohm version$^{(\Bohm)}$ of the EPR {Gedankenexperiment},
which violates the Bell inequality as a result of the coincidence counting.
Quantum mechanics describes statistical results of measurements economically,
but it does not explain how the results occur.
To see clearly whether there exists action at a distance or not, we have to
refer to states of the objects before and after measurements, i.e., processes
of measurements.
Hence our model comprises hidden-variables.
The purpose of this paper is neither to explain why the values of spin are
quantized nor to replace quantum mechanics by classical mechanics.
The aim of this paper is to find an example, at least in thinking, that shows
that the coincidence counting is the source of the non-locality.
In our model, in order to describe the time evolution of a measuring process,
we use a system of differential equations which has attractors.
Since the attractors are invariant for the flow, they are invariant for the
measurement.
So the attractors are related to corresponding quantum-mechanical eigenstates.
This is a new feature of our model.

The organization of this paper is as follows:
In $\S$2, we shall construct a model of a measurement for a single spin-1/2
object.
This section contains also preparation of the next section.
In $\S$3, we shall make an extension of the model of $\S$2 to the EPR-Bohm
situation.
This section is the heart of this paper.
By taking an appropriate closing time in sampling data, our model reproduces
the correlations predicted by quantum mechanics approximately.
Section 4 is devoted to discussion and summary.

\chapter{ A hidden-variable model of a measurement of a single spin-1/2 object}
In this section, we shall exhibit a model of  measuring processes of a spin-1/2
object.
A measuring apparatus changes a state of the object due to interaction between
them, hence this change is not instantaneous generally.
Therefore a system of ordinary differential equations is used in order to
describe this change.
For a different setting of the measuring apparatus, time evolution of a state
of the object is governed by  a different system of differential equations.
For different settings of the measuring apparatus, the probability spaces of
outcomes are different .
Thus our model becomes a contextual hidden-variable theory.

A spin-1/2 object is not a mere point particle but a system of many degrees of
freedom, because the spin can be considered as degrees of freedom that describe
a rotation of the object on some axis.
We denote these degrees of freedom by $\Spin=(S_x, S_y, S_z) \in \Real^3$.
Quantum mechanics gives us the following information about the spin:
For convenience, using the unit such that $\hbar$=1, we put $j=1/2$,
$J=\sqrt{3/4}$ and denote the quantum-mechanical observables of the spin by
three-tuple of operators $(\hat{S_x},\hat{S_y},\hat{S_z})$;
The up(down)-eigenstate of $\hat{S_z}$ has the eigenvalue $+j$($-j$) for
$\hat{S_z}$ and $J^2$ for $\hat{S_x}^2+\hat{S_y}^2+\hat{S_z}^2$, respectively.
Hence the up-eigenstate of $\hat{S_z}$ may correspond to an ensemble whose
members have the properties of $S_z=+j$ and $|\Spin|=J$.
$\Spin$ is yet insufficient to describe a state of the spin-1/2 object, since
$\Spin$ itself that satisfies above properties is not parallel to the $z$-axis.
We must take account of other degrees of freedom that express that a state is a
member of an ensemble corresponding to the up-eigenstate.
We denote them by ${\bf U}=(U_x, U_y, U_z)$.
In actual experiments, it does not matter when the object enters the measuring
apparatus and when it escapes from it.
Hence we do not have to take  account of details of the motion of the object in
the actual space.
However they become important for  coincidence counting.
And we shall take them into account in the next section.
We shall see in this section that the six degrees of freedom are sufficient to
model the measuring process of a single spin-1/2 object.

We denote a six dimensional space $\Real^6$ whose coordinates are given by the
$(\Spin, \hidV)=(S_x, S_y, S_z, U_x, U_y, U_z)$ by $\PS$.
A state of the object is represented by a point $(\Spin, \hidV)$ in $\PS$.

Let $\alpha_1$ be a subset of $\PS$ that is defined as
$$
\alpha_1= \{ (\Spin, \hidV)\in \PS : S_z=+j, |\Spin|^2=J^2, U_x=U_y=U_z-J=0
\}.$$
As stated above, we identify an ensemble of states distributed uniformly in
$\alpha_1$ with the up-eigenstate of $\hat{S_z}$.
In the same way, let $\alpha_2$ be a subset of $\PS$ that is defined as
$$
\alpha_2= \{ (\Spin, \hidV)\in \PS : S_z=-j, |\Spin|^2=J^2, U_x=U_y=U_z+J=0
\}$$
and  we identify an ensemble of states distributed uniformly in $\alpha_2$ with
the down-eigenstate of $\hat{S_z}$.

A good measuring apparatus for $z$-component of the spin must effect such a
change of a state of the object that the state  approaches to either $\alpha_1$
or $\alpha_2$.
Therefore the measuring process of $z$-component of the spin may be modeled by
a system of differential equations for which $\alpha_1$ and $\alpha_2$ are
attractors$^{(\ATTR)}$.
The simplest one among such systems of differential equations may be the
following:
$$ \left\{ \eqalign{
{d\Spin \over dt}=& \hidV\times\Spin-\epsilon_1 P_{xy} {\partial \psi \over
\partial \Spin} \psi-\epsilon_1 \bigl\{ \theta(S_z-\beta)\phi_+ +
\theta(-S_z+\beta)\phi_- \bigr\} \ez, \cr
{d\hidV \over dt}=& -\epsilon_2 P_{xy} \hidV -\epsilon_2 \bigl\{
U_z-\epsilon(S_z-\beta)J \bigr\} \ez -\epsilon_2 U_z \bigl\{ |\hidV|^2-J^2
\bigr\} \ez, \cr
} \right.      \eqn\DiffEq
$$
where $\epsilon_1=10.0$, $\epsilon_2=0.05$ and
$$\eqalign{
\psi(\Spin)\equiv & |\Spin|^2-J^2,\cr
\phi_\pm(\Spin)\equiv & S_z \mp j, \cr
\omega(\hidV)\equiv & \cos^{-1}(U_z/|\hidV|),\cr
\beta(\omega) \equiv & \bigl\{ j \cos\omega -\sqrt{J^2-j^2}\cos({\pi\over
2}(1-\cos\omega)) \sin\omega \bigr\} \cr
&\qquad \times \bigl\{ 0.98 \theta(|\cos \omega |-0.99)+\theta(0.99-|\cos
\omega|) \bigr\},\cr
\ez=&\left( \matrix{ 0 \cr 0 \cr 1 \cr}\right),
\qquad  P_{xy}=\left(\matrix{1 & 0 & 0 \cr
                                0 & 1 & 0 \cr
                                0 & 0 & 0 \cr}\right).\cr
}
$$
$\theta(x)$ is the step function that is defined as $\theta(x)=1$, if $x \ge
0$; $\theta(x)=0$, otherwise.
And $\epsilon(x)$ is the sign function that is defined as $\epsilon(x)=1$, if
$x \ge 0$; $\epsilon(x)=-1$, otherwise.
For simplicity, the unit of time is chosen appropriately.
Thus we assume that the evolution of a state $(\Spin, \hidV)$  during {\it the
measurement of $S_z$}, i.e., $z$-component of the spin, is governed by
$\DiffEq$.

We wish to make  some remarks on Eq.$\DiffEq$.
The terms containing $\epsilon_i$'s, $i=1,2$, of Eq.$\DiffEq$ are crucial for
the existence of the attractors $\alpha_1$ and $\alpha_2$.
To see this, suppose that there is no terms in the right-hand sides of
Eq.$\DiffEq$ except for $-\epsilon_1 P_{xy} ({\partial \psi / \partial \Spin})
\psi $.
Then $(d\psi^2/dt)$=$2(\partial \psi/\partial \Spin)\cdot(d
\Spin/dt)\psi$=$-8\epsilon_1 (S_x^2+S_y^2)\psi^2\leq0$.
If neither $S_x^2+S_y^2$ nor $\psi$ vanish, then $\psi(\Spin(t))^2$ is strictly
monotone decreasing as a function of time $t$.
As $t\rightarrow \infty$, $\psi^2$ may vanish, i.e., $|\Spin|^2$ may approach
to $J^2$.
It is similar for the other terms containing $\epsilon_1$ or $\epsilon_2$.
The role of the function $\beta(\omega)$ is to give a border of $S_z\rightarrow
+j$ or $S_z\rightarrow -j$.
Thus the terms containing $\epsilon_i$'s, $i=1, 2$, represent effectively  the
influence on a state of the object of the interaction with the measuring
apparatus.

In actual experiments, we use, for example, the Stern-Gerlach magnet as a
measuring apparatus of $S_z$.
In this case, the $z$-component of spin is not directly measured.
We judge $S_z=+j$ or $-j$ according to the sign of $z$-component of the
velocity of the object gained eventually by the non-uniform magnetic field.
Since $\Spin$ may behave as if a magnetic dipole, if $\Spin$ is stabilized in a
neighborhood of $\alpha_1$($\alpha_2$), then $S_z>0$($<0$), so the object gains
positive(negative) $z$-component of velocity.
{}From these considerations, let us regard the measurement of $S_z$ in our
model as the following procedure:
A state $(\Spin, \hidV)$ of the object begins to evolve by the equation
$\DiffEq$, when the interaction between the object and the measuring apparatus
is switched on.
When the state in $\PS$ comes into an appropriate neighborhood $G(\alpha_1)
\equiv \{ (\Spin, \hidV)\in \PS : |S_z-j|<\delta \}$ of $\alpha_1$,
$\delta=0.01$,  the measurement finishes and we obtain the outcome $+j$.
Otherwise, when the state in $\PS$ comes into an appropriate neighborhood
$G(\alpha_2) \equiv \{ (\Spin, \hidV)\in \PS : |S_z+j|<\delta \}$ of
$\alpha_2$, the measurement finishes and we obtain the outcome $-j$.

Rigorously speaking, the measured $z$-component of the spin is not represented
by $S_z$; rather it is represented by a slightly modified function ${\tilde
S}_z$ defined on $\PS$ as
$$
{\tilde S}_z(\rho)=\cases{ +j, & if $\rho\in G(\alpha_1)$, \cr
                       -j, & if $\rho\in G(\alpha_2)$, \cr
                       S_z(\rho), & otherwise. \cr}
$$
For brevity, we shall write that $S_z\Rightarrow j$ in place of ${\tilde
S}_z=j$ and so on hereafter.
We have finished the presentation of the model of a measuring process of $S_z$
here.

Let $\Rot$ be a rotation (matrix) by an angle $\theta$ along an axis in $x$-$y$
plane in $\Real^3$.
When we perform a measurement of $\Spin\cdot\Rot\ez$, the evolution of a state
is given by a system of differential equations that is obtained by rotating
Eq.$\DiffEq$.
Let $\beta_\theta$ be a subset of $\PS$ defined as $\beta_\theta \equiv
\{(\Rot\Spin, \Rot\hidV)\in \PS : (\Spin, \hidV)\in \alpha_1  \}$, i.e., it is
obtained by rotating $\alpha_1$.
Then the $\beta_\theta$ is an attractor associated with the property of
$\Spin\cdot\Rot\ez\Rightarrow j$.

As an initial condition, we study an ensemble of states distributed uniformly
in the $\beta_\theta$.
When we perform the measurements of $\Spin\cdot\Rot\ez$ for this ensemble,
$\Spin\cdot\Rot\ez \Rightarrow j$ holds with certainty, since all the members
of the ensemble are already in  the neighborhood  $G({\beta_\theta})$ of the
attractor $\beta_\theta$.
Hence such an ensemble can be identified with an eigenstate of the
quantum-mechanical observable corresponding to $\Spin\cdot\Rot\ez$.
Taking this ensemble as the initial condition and solving Eq.$\DiffEq$
numerically, we have calculated the probabilities that the trajectories come
into the neighborhoods $G({\alpha_1})$ of $\alpha_1$ and $G({\alpha_2})$ of
$\alpha_2$, respectively.
We show in Fig. 1 the probabilities of outcomes of the measurements of
$S_z=\Spin\cdot\ez$  being $+j$ with respect to the ensembles of states
distributed uniformly in $\beta_\theta$  for several relative angles $\theta$.
The result fits  with the prediction of quantum mechanics.
In this sense, we can say that the model of this section is the one of
measuring processes of a single spin-1/2 object.

Before closing this section, we wish to make a remark.
We call the time when the interaction between the object and the measuring
apparatus is switched on  a {\it beginning time}.
And we call the time when the state comes into one of the neighborhoods of the
two attractors  a {\it finishing time}.
Since we may put the beginning time for each states to be zero without loss of
generality, we do so hereafter.
The finishing time for each objects depends both on what is measured and on the
initial states.
In general, finishing times fluctuate due to the variation of initial states,
even if the beginning times are the same.
This fluctuation plays an important role when we use the coincidence counting
as in the next section.

\chapter{A local model of the EPR-Bohm {Gedankenexperiment}}

Now we shall make an extension of the previous model to the EPR-Bohm
Gedankenexperiment.
We consider two spin-1/2 objects which are distinguished by labeling with $A$
and $B$.
Let $\PS_A$ and $\PS_B$ be the phase spaces of the spin-1/2 objects $A$ and
$B$, respectively.
The whole phase space $\PS_{AB}$ is given by the direct product
$\PS_A\times\PS_B$.
Suppose that two measuring apparatuses are placed apart on the $y$-axis at
equidistance from the origin.
For convenience, we shall measure components of the spins along  directions
perpendicular to the $y$-axis.

The EPR-Bohm quantum-mechanical state is a singlet state.
We note that it is rotationally invariant;
and in the quantum-mechanical state, the pair of spins has completely negative
correlations.
Let $s$ be a subset of $\PS_{AB}$ defined as
$$\eqalign{
s \equiv \bigcup_{\phi=0}^{2 \pi}\bigcup_{\theta=0}^{\pi} \biggl\{ & (\Spin_A,
\hidV_A, \Spin_B, \hidV_B)\in \PS_{AB} : \Spin_A+\Spin_B={\bf 0},
\hidV_A+\hidV_B={\bf 0}, \cr
  & \qquad \bigl(\Roty(-\theta)\Rotz(-\phi)\Spin_A,
\Roty(-\theta)\Rotz(-\phi)\hidV_A\bigr)\in \alpha_1 \biggr\}, \cr
}$$
where $\Rotz(\phi)$ and $\Roty(\theta)$ represent the rotation along the
$z$-axis by an angle $\phi$ and the rotation along the $y$-axis by an angle
$\theta$, respectively.
An ensemble of states of pairs  distributed uniformly in the subset $s$ has the
above mentioned two features of the EPR-Bohm quantum-mechanical state.
Hence let us take this ensemble as an initial condition just before
measurements in the EPR-Bohm situation.

In the actual experiment, the coincidence counting is used in order to identify
detected objects as a pair.
Therefore the time when the object escapes from the measuring apparatus is
important.
Since the actual motion of the object in the measuring apparatus may be
complicated, in stead of modeling the details of the motion concretely, we just
assume that there exists a threshold time $T$ such that if the finishing time
is greater than $T$, then as a result of the interaction with the measuring
apparatus, the value of $y$-coordinate of the object becomes random.
We can also rephrase this assumption as follows:
There is $T$ such that if the finishing time is greater than $T$, then the time
when the object escapes from the measuring apparatus fluctuates.
We emphasize that this assumption concerns itself with only each of measuring
apparatuses and the objects, so no action  at a distance is stolen into our
model by this assumption.
It shall be shown later that by the coincidence counting, a pair of spin-1/2
objects is taken into account as outcomes of measurements, only if their
finishing times are less then $T$.
Hence we call $T$ a {\it closing time} in sampling data.

We calculated  correlations of the spins of the objects $A$ and $B$ with a
closing time $T$= 0.133 numerically as in the previous section.
The results are plotted in Fig. 2 and compared with the quantum-mechanical
correlations.
Our model approximates the quantum-mechanical correlations.
For comparison, we also calculated the correlations without the closing time,
i.e., without using the coincidence counting.
The results are plotted in Fig. 3.
In this case, since we instituted no closing time, the correlations were
calculated in a single probability space independent of what are measured.
Since our model has no action at a distance, as expected from the
no-go-theorems$^{(\NoGo)}$ of non-contextual hidden-variable theories, the
result does not agree with quantum mechanics without the closing time.

Figure 4 represents the results of calculations of the quantity $F(\phi)$ that
appears in Ref. $(\Redh)$.
The Bell inequality implies that $F(\phi)$ does not exceed two.
The results with the closing time $T=0.133$ agree  with quantum mechanics
approximately and the Bell inequality is violated.
The results without the closing time, on the other  hand, satisfy the Bell
inequality and do not agree with quantum mechanics.

Our task is now to express the assumption for the motion of the position of the
object more concretely and show that the coincidence counting leads to the
institution of a closing time in sampling data.
Suppose that the devices of the Gedankenexperiment are arranged as follows (see
Fig. 5):
The source of pairs of the objects is at the origin.
Measuring apparatuses for the object $A$ and the object $B$ are placed apart on
the $y$-axis at equidistance from the origin.
Let $W$ be their lengths along the $y$-direction.
A detector for the object $A$ is placed behind the measuring apparatus, say on
$(0, -L ,0)$, viz., on the $y$-axis at the distance $L$ from the origin to the
negative direction.
In the same way, a detector for the object $B$ is placed on $(0,L,0)$.

Let $t_0$ be the earliest time when the objects reach the each detectors.
Recall here that all the beginning times are zero.
Let $v_0$  be modulus of $y$-component of the velocity of each object at the
outside of the measuring apparatus.
For the object such that its finishing time is less than $T$, we denote the
modulus of $y$-component of the velocity at the inside of the measuring
apparatus by $v$(=const.).

Then our assumption can be expressed clearly as follows:
For an object with its finishing time $\tau$, let us assume that if $\tau \leq
T$, then the time when the object escapes from the measuring apparatus is
$W/v$;
 otherwise, the time when the object escapes from the measuring apparatus
fluctuates uniformly in a time interval $[W/v, (W/v)+\tau]$ as a result of the
interaction between the object and the measuring apparatus.
Thus the closing time $T$ characterizes the way of diffusion of the position of
the object in the measuring process.

Let $\TA$ be a finishing time  for a spin-1/2 object $A$.
Then the probability of the object $A$ existing in an interval $[y, y+\Dy]$ of
width $\Dy \ll 1$  on the $y$-axis at the time $t_0$ becomes approximately
$$
\rho_A(y)\Dy = \theta(T-\TA)\chi_{[-L, -L+\Dy]}(y)+\theta(\TA-T) {1\over v_0
\TA} \chi_{[-L, -L+v_0 \TA]}(y)\Dy.
$$
Here for a subset $E$ of the real line, $\chi_E$ represents the
characteristic(defining) function of $E$ that is defined as $\chi_E(\rho)=1$,
if $\rho\in E$; $\chi_E(\rho)=0$, otherwise.

In the same way,
let $\TB$ be a finishing time for a spin-1/2 object $B$.
The probability of detection of the object $B$ in an interval $[y, y+\Dy]$ of
width $\Dy \ll 1$ at the time $t_0$ becomes approximately
$$
\rho_B(y)\Dy = \theta(T-\TB)\chi_{[L-\Dy, L]}(y)+\theta(\TB-T) {1\over v_0 \TB}
\chi_{[L-v_0\TB, L]}(y)\Dy.
$$

For the objects $A$ and $B$ whose finishing times are $\TA$ and $\TB$,
respectively, we estimate the probability $p_c$ of coincidence detection.
We partition the $y$-axis into intervals $[y_n, y_{n+1})$, $n\in{\bf Z}$, of
width $\Dy$, where $y_n=n \Dy$.
Since the coincidence detection is done not only at the time $t_0$ but also at
delayed times, it becomes, for $\Dy \ll 1$,
$$\eqalign{
p_c =& \sum_n \rho_A(-y_n)\Dy \rho_B(y_n)\Dy \cr
\approx& \int dy \rho_A(-y) \rho_B(y) \Dy \cr
=& \theta(T-\TA)\theta(T-\TB)+O( \Dy) \cr
=& \theta(T-\TA \vee \TB)+O( \Dy), \cr
}
$$
where $\TA\vee\TB$ represents the maximum value between $\TA$ and $\TB$.
Thus if the accuracy $\Dy$ of position is very small, then the behavior of the
$p_c$ is like $\theta(T-\TA\vee\TB)$.
This means that if at least one of the finishing times $\TA$ and $\TB$ is
greater than $T$, then the probability of the objects $A$ and $B$ being
detected at the same time vanish.
Therefore the detected objects $A$ and $B$ at coincidence have  finishing times
$\TA$ and $\TB$ both of which are less than $T$  with certainty.

Accordingly, as far as we use the coincidence counting, our local model
violates the Bell inequality.

\chapter{Discussion and Summary}
We constructed a local model of spin-1/2 objects in the EPR-Bohm Gedanken-
experiment.
The fact that there exists a local model that violate the Bell inequality even
for spin-1/2 objects supports the idea that the coincidence counting is the
source of the non-locality and there exists no action at a distance in the
EPR-Bohm situation.

Our model is an example that shows that interaction between the object and the
measuring apparatus is not the unique reason why the probability space of
outcomes of measurements changes according to what are measured.
In fact, by introducing the closing time $T$, our model of $\S$3 produces
different probability spaces of outcomes according to the choice of different
settings of the measuring apparatus.
Thus this local model becomes a contextual hidden-variable theory.
In our model, the sample space changes according to the choice of $\theta_{\bf
ab}$, the angle between the directions ${\bf a}$ and ${\bf b}$ of the spins of
the objects $A$ and $B$ to be measured.
Nevertheless, as shown in Fig. 6, the change of the number of samples is small
(less than 10 \%).
Our results seem to be consistent with the results of the experiment$^{(\Asp)}$
of Aspect {\it et al.}, though it was done for photons, in which there is no
change in the number of samples.

In $\S$2, we identified the ensemble of the states of the object distributed
uniformly in each attractors with the corresponding eigenstates.
In $\S$3, we have identified the ensemble distributed in the subset $s$ of
$\PS_{AB}$ with the EPR-Bohm quantum-mechanical state.
One may ask what ensemble corresponds to a given quantum-mechanical state.
The question is beyond the scope of this paper, because it is almost equivalent
to understand the superposition principle of quantum states.
In fact, in order to find the answer, we must understand the meaning of a phase
of a quantum-mechanical state vector.
For a stationary quantum-mechanical state vector, the phase evolves in time.
This suggests that the phase has information of the dynamical evolution of the
state of the object before the beginning of a measurement.
Our phenomenological model lacks this information, since it treats only the
dynamics after the beginning of a measurement.

However, it is interesting and important to find the answer, because this has
deep connection with whether our human reason can understand things exist in
the external world or not.
It is known that propositions for quantum phenomena are subject to some
non-Boolean logic$^{(\Watanabe)}$.
These propositions are concerned  with outcomes of measurements.
While, since the human reason is subject to the Boolean logic, the outcomes of
the measurements contradict the human reason.
Accordingly, in order to understand the things behind the quantum phenomena by
the human reason, we cannot help assuming something that is subject to the
Boolean logic.
Thus to understand the things in the external world means necessarily
introduction of some hidden-variables into theory.
Further, we must comprehend how the Boolean object characterized by the
hidden-variables produces such non-Boolean phenomena as quantum phenomena.
Although our model in this paper is  so restricted that it may have less
connection with the things existing in the external world, it gives an example
such that a Boolean object leads to  non-Boolean phenomena.
In this sense, our model is instructive.
Our model suggests that the things existing in the external world would be
local, too.

We summarize our results in the following:
We have constructed a local hidden-variable model of spin-1/2 objects in the
EPR-Bohm { Gedankenexperiment}.
By instituting the appropriate closing time in sampling data, the correlation
that is calculated by our model approximates the quantum-mechanical
correlation.
{}From this, as far as we use the coincidence counting, our local model
violates the Bell inequality with no action at a distance.

\ack
I wish to express my gratitude to Professor T. Ishigaki and Professor K.
Ishikawa for many helpful discussion and valuable suggestions and to Professor
K. Fujii for interesting discussion.

\refout
\endpage
\figout
\end